%% file: Main_text_incl._figures.tex
\documentclass[reprint,aps,pre,showpacs,superscriptaddress,groupedaddress,twocolumn]{revtex4-1} 
\usepackage{graphicx,floatrow,subfig,amsmath,amssymb,dcolumn,bm,chngcntr,epstopdf,tikz}

\DeclareMathOperator\arctanh{arctanh}

\input{defs}

\begin{document}
\preprint{APS/123-QED}
\title{Fabry-Perot resonance of water waves}
\author{Louis-Alexandre Couston,$^1$ Qiuchen Guo,$^1$ Maysamreza Chamanzar,$^2$ and Mohammad-Reza Alam$^1$}
\affiliation{$^1$ Department of Mechanical Engineering, University of California, Berkeley, CA 94720, USA\\
$^2$ Department of Electrical Engineering \& Computer Sciences, University of California, Berkeley, CA 94720, USA}

\begin{abstract}

We show that significant water wave amplification is obtained in a water resonator consisting of two spatially separated patches of small-amplitude sinusoidal corrugations on an otherwise flat seabed. The corrugations reflect the incident waves according to the so-called Bragg reflection mechanism, and the distance between the two sets controls whether the trapped reflected waves experience constructive or destructive interference within the resonator. The resulting amplification or suppression is enhanced with increasing number of ripples, and is most effective for specific resonator lengths and at the Bragg frequency, which is determined by the corrugation period. Our analysis draws on the analogous mechanism that occurs between two partially reflecting mirrors in optics, a phenomenon named after its discoverers Charles Fabry and Alfred Perot.

\end{abstract}

\pacs{}

\maketitle

\counterwithout{equation}{section}

\allowdisplaybreaks

\section{Introduction}

Fabry-Perot cavities are standing-wave resonators commonly used in optics, quantum physics, and astronomy \cite{Fabry1897,Bland1989,Vahala2003}. In its simplest form, an optical  Fabry-Perot cavity consists of two partially reflecting mirrors surrounding a dielectric medium. Light waves entering the cavity undergo multiple partial reflections between the two mirrors which constructively interfere at resonance frequencies determined by the round trip propagation delay and the phase shifts incurred at the mirrors \cite{Vaughan1989}. The Fabry-Perot device was originally applied in interferometry, but is now also used in laser resonators due to its ability in amplifying the radiation field within the cavity \cite{Hernandez1988}.

Seafloor variations in the ocean can, much like mirrors and lenses in optics, significantly affect the propagation of incident waves. While seabed inhomogeneities generally lead to water wave scattering due to the absence of coherence between the multiple scattered waves, instances of constructive interference due to periodic undulations of the seabed have been observed in nature, such as in the Rotterdam waterway \cite{Kranenburg1991,Pietrzak2004}, Cape Cod Bay in Massachusetts \cite{Elgar2003}, and near numerous shorelines \cite[][]{Mei2001}. The strong reflection of surface waves by bottom corrugations, which has also been demonstrated in the laboratory \cite{Heathershaw1982,Benjamin1987,Hara1987}, relies on the well-known Bragg mechanism discovered in solid-state physics \cite{Bragg1913} and first reported in the context of water waves by Davies \cite{Davies1982}: surface waves with wavelengths twice the wavelength of seabed corrugations experience coherent reflections.  Interestingly, Bragg resonance between surface waves and seabed corrugations is also the reason why natural sandbars, which can be seen parallel to shore in many coastal areas \cite[][]{Mei2001}, are sinusoidal with wavelength equal to half that of the local surface waves \cite[][]{Rey1995,Yu2000b}.  The same Bragg reflection is found in optics when light waves encounter multilayer dieletric coatings that offer significant advantages over single-layer mirrors \cite{Hernandez1988}. 

Even though seabed corrugations act like partially reflective mirrors, the Bragg reflection of water waves does not always lead to decreased wave activity downstream of the corrugations. Indeed,  Yu \& Mei \cite{Yu2000} confirmed the earlier conjecture \cite{Kirby1990} that the presence of a bar patch upstream of a reflective beach could result in shoreward wave amplification, rather than attenuation, for specific patch-to-shore distances. Their result has now been extended to the normal modes of oscillation of a corrugated wave tank \cite{Howard2007,Weidman2015} and will be related to the resonance studied in this work.

\begin{figure}
\centering\includegraphics[width =3.4in]{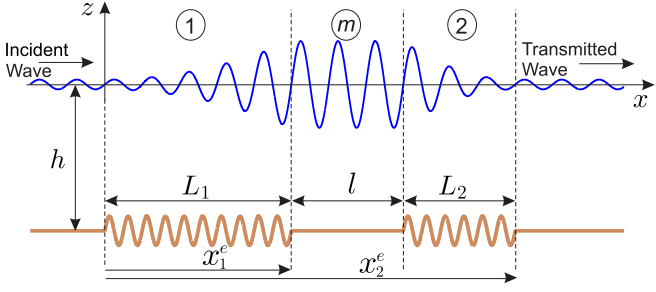}
\caption{(Color online) Schematic of a water wave Fabry-Perot resonator. Surface waves interact with two patches of ripples (wavelength $\lambda_b$) on an otherwise flat seabed. $l$ is the resonator length and $L_{1,2}=N_{1,2}\lambda_b$, $N_{1,2}\in\mathbb{N}$.}\label{setup}
\end{figure}

Here we show that significant wave amplification or suppression can be achieved in a region of constant water depth bounded by two sets of small-amplitude corrugations (see Fig. \ref{setup}). We thus demonstrate the analogy between the underlying water wave trapping mechanism and the Fabry-Perot resonance in optics based on distributed Bragg mirrors \cite[][]{Mangaiyarkarasi2006,Numai2015}. We obtain the resonance condition for water wave amplification and suppression close to the Bragg frequency using multiple-scale analysis, and we investigate the effect of the reflectivity of the patches as well as resonator length between the two mirrors on the field enhancement and transmission spectra. Our results are obtained within the framework of the linear potential flow theory, and are then extended in the conclusion by discussing and providing suggestions on how to consider the effects of wave directionality, bottom irregularity, and viscosity.   \\

\section{Bragg mirrors for water waves}

Consider the propagation of surface gravity waves on an incompressible, homogeneous, and inviscid fluid. The flow is assumed irrotational such that the velocity field $\textbf{u}$ can be expressed in terms of a velocity potential $\phi$ as $\textbf{u}=\nabla \phi$. In a Cartesian coordinate system with $x,y$ axis on the mean free surface and $z$ axis positive upward, the linear form of the governing equations for $\phi$ reads
\bsa{100}
&\nabla^2\phi=0,\hspace{2.5cm}  -h<z<0\\
&\phi_{tt}+g\phi_z=0, 
\hspace{2.5cm}  z=0,\\ \label{00}
&\phi_z=\nabla_H\zeta\cdot \nabla_H\phi, \hspace{1.5cm}  z=-h+\zeta(x,y),
\esa
in which $\nabla_H=(\p_x,\p_y)$ is the horizontal gradient operator, $z=-h$ describes the position of the mean seabed, and $\zeta$ represents the small seabed undulations. The free surface elevation $\eta(x,y,t)$ is related to the velocity potential through $\eta=-\phi_t/g$. 

 Here, we assume that the problem is two-dimensional, i.e., $\p/\p y \equiv 0$. On a flat seabed (i.e. $\zeta=0$), the left- and right-propagating wave solutions of Eqs. \eqref{100} have constant amplitudes and the free-surface elevation reads
\ba{}
\eta(x,t) = \lp \mathcal{A}\e^{-ikx} + \mathcal{B}\e^{ikx}\rp\e^{i\omega t} + \text{c.c.}
\ea  
with c.c. denoting the complex conjugate. The surface wavenumber $k$ and the wave frequency $\omega$ are related through the dispersion relation
\ba{}\label{disp}
\omega^2=gk\tanh kh.
\ea 
Above a region with corrugated seafloor (e.g., no. 1 or no. 2  in Fig. \ref{setup}), the waves have varying amplitudes due to wave-seabed interactions, and the general solution becomes
\ba\label{101}
\eta(x,t) = \lb \mathcal{A}(x,t)\e^{-ikx} + \mathcal{B}(x,t)\e^{ikx}\rb\e^{i\omega t} + \text{c.c.}
\ea
The classical case of a single patch with corrugations of the form
\ba\label{103}
\zeta(x) = \lcb  \begin{array}{c} 
d\sin\lb k_b (x-x^s) - \theta \rb, \\
0, 
\end{array} \hspace{0cm} \begin{array}{c} 
x\in[x^s,x^e], \\
\text{elsewhere},
\end{array}  \right. 
\ea 
where $d$ is the amplitude of the ripples, $\theta$ the corrugations' phase, and $x^s$ and $x^e$ the start and end of the patch, is known to strongly reflect surface waves with wavenumber $k=k_b/2+\kappa$, $\kappa/k_b \ll 1$. For waves coming from $x=-\infty$ and small corrugation amplitude $k_b d\ll 1$, Mei \cite{Mei1985} showed that the steady-state solution for the wave envelope amplitudes leads to the so-called Bragg reflection and transmission coefficients here rewritten as
\bsa{2011} \label{2011a}
\mathcal{R}^B = \f{\mathcal{B}(x^s)}{\mathcal{A}(x^s)} = \f{\e^{-i\theta} \sinh S w }{ w \cosh S w  + i \varpi\sinh S w  }, \\
\mathcal{T}^B = \f{\mathcal{A}(x^e)}{\mathcal{A}(x^s)} = \f{w}{ w \cosh S w  + i \varpi\sinh S w  },
\esa
where
\bsa{2012} \label{2012a}
&\varpi=\Omega/\Omega^c,~w = \sqrt{1-\varpi^2},~\Omega^c=\f{\omega_b k_b d}{4\sinh k_bh},   \\  \label{2012b}
& S = \f{\Omega^c L}{C_g} =  \f{N\pi  k_b d }{2(\sinh k_b h + k_b h)}.
\esa
The reflection and transmission coefficients \eqref{2011} are valid for wave frequencies in the vicinity of the Bragg frequency, i.e., for $\omega=\omega_b+\Omega$ where $\omega_b=\omega(k=k_b/2)$ and $\Omega=C_g\kappa \ll \omega_b$ with $C_g=d\omega/\d k$ the wave group velocity [cf. Eq. \eqref{disp}]. In Eqs. \eqref{2012}, $N$ is the number of corrugations, and the parameter $\Omega^c$ used to normalize the dimensional detuning frequency $\Omega$ is called the cut-off frequency since for $\Omega<\Omega^c$ the envelope modulations are exponential over the corrugations whereas they are oscillatory for $\Omega>\Omega^c$. The variable $S$, which can be rewritten as $\tanh S=|\mathcal{R}^B(\varpi=0)|=R^B_0$, is a measure of the  reflected wave amplitude at the Bragg frequency.

\begin{figure}
\centering\includegraphics[width =3.0in]{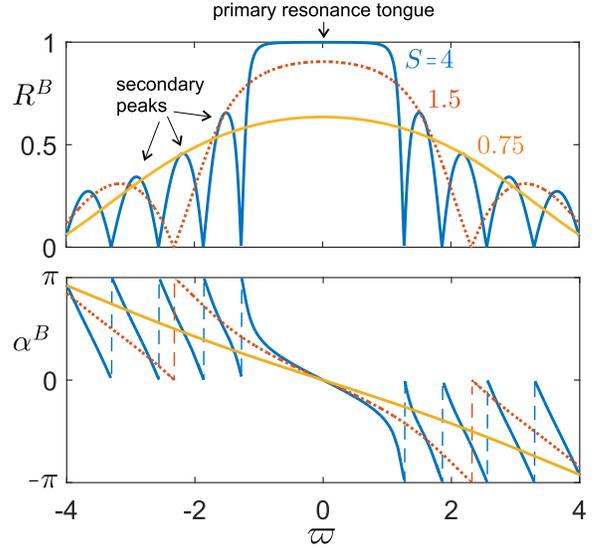}
\caption{(Color online) Effect of detuning $\varpi$ on the normalized reflected wave amplitude $R^B=|\mathcal{R}^B|$ and phase $\alpha^B=\arg(\mathcal{R}^B)$ off a water Bragg reflector with $\theta=0$ [cf. Eq. \eqref{2011a}]. $R^B$ and $\alpha^B$ are shown for $S=0.75$ (solid), 1.5 (dotted), 4 (solid) while keeping the cutoff frequency $\Omega^c$ fixed [cf. Eq. \eqref{2012a}]. The reflection strength $R^B$ is strongest and almost constant within the primary resonance tongue, unlike $\alpha^B$. }\label{reflect}
\end{figure}

The effect of detuning on the reflected wave amplitude $R^B=| \mathcal{R}^B |$ and phase shift $\alpha^B = \arg (\mathcal{R}^B)$ off a single patch of corrugations is shown in Fig. \ref{reflect} for various values of the parameter $S$. As expected, $R^B_0$ increases with $S$, i.e., with an increase in the number of ripples $N$ or ripple amplitude $d/h$, as well as with a decrease in the normalized water depth $k_bh$. While increasing $d/h$ is analogous to increasing the refractive index contrast in mirrors made of alternating dielectrics in optics \cite{Hernandez1988}, here we remark that the effect of decreasing $k_bh$ on the reflection coefficient has no direct equivalent in optical systems. This is of course due to the fact that water waves are surface waves, which experience stronger seabed effects with smaller normalized water depth $kh$. Water-wave dispersion by the fluid medium yet vanishes in the long-wave regime, i.e., when $kh \ll 1$, in which case $\mathcal{R}^B$ becomes independent of $k_bh$ and reaches a maximum value.  The primary resonance tongue, i.e., where $R^B$ is the strongest and only mildly varying, typically extends to $|\varpi|\leq 1$. At the Bragg frequency, $\alpha^B = -\theta$ such that a set of ripples acts like a partially reflecting wall when $\theta=0$, with a free-surface anti-node formed at the beginning of the patch. Unlike $R^B$, the phase shift $\alpha^B$ changes significantly within the primary resonance tongue (see Fig. \ref{reflect}). Indeed, $\alpha^B$ increases for longer wavelengths (i.e., $\varpi<0$), which corresponds to a downstream displacement of the anti-node upwave of the leading seabed crests. The phase shift due to detuning extends up to $\pm \pi/2$ at the edges of the primary resonance tongue when $S\rightarrow\infty$.  \\

\section{Fabry-Perot resonance}

Let us now construct the water wave analog of an optical Fabry-Perot cavity using two patches of seabed corrugations as water wave mirrors. We consider the seafloor variations 
\ba\label{1003}
\zeta(x) = \lcb  \begin{array}{c} 
d\sin\lb k_b (x-x^s_1) - \theta_1 \rb, \\
d\sin\lb k_b (x-x^s_2) - \theta_2 \rb, \\
0, 
\end{array} \hspace{0cm} \begin{array}{c} 
x\in[x^s_1,x^e_1], \\
x\in[x^s_2,x^e_2], \\
\text{elsewhere},
\end{array}  \right. 
\ea 
where $x^s_1=0,~x^s_2=x^e_1+l ,~x^e_{1,2}-x^s_{1,2}=N_{1,2}\lambda_b$  (see Fig. \ref{setup}). Subscripts 1 and 2 apply to variables for the upstream and downstream patch respectively. We refer to the parameter $l$ separating the two patches as the resonator length. Similarly to the Bragg reflection and transmission strength coefficients obtained for each one of the two patches taken individually, noted $R^B_{1,2}$ and $T^B_{1,2}$, we define a Fabry-Perot reflection  and transmission strength coefficient as $R^{FP} = |\mathcal{B}(x^s_1)/\mathcal{A}(x^s_1)|$ and $T^{FP}=|\mathcal{A}(x^e_2)/\mathcal{A}(x^s_1)|$. The derivation (provided in the Appendix) yields
\bsa{204}
&R^{FP} = \lb \f{(R_1^B)^2+(R_2^B)^2-2 R_1^B R_2^B \cos \gamma}{1+(R_1^BR_2^B)^2-2 R_1^B R_2^B \cos \gamma} \rb^{1/2}, \\
&T^{FP} =  \lcb \f{\lb 1-(R_1^B)^2 \rb\lb 1-(R_2^B)^2 \rb}{1+(R_1^BR_2^B)^2-2 R_1^B R_2^B \cos \gamma} \rcb^{1/2},
\esa
where 
\ba{}\label{205}
\gamma = \pi-2\theta_1+2kl-\alpha_1-\alpha_2
\ea
is the round trip phase shift. In Eq. \eqref{205}, $2kl$ is the propagation phase accumulation and $\alpha_{1,2}$ are the phase shifts incurred upon reflection at the Bragg mirrors. When $\gamma=2m\pi$ ($m\in\mathbb{N}$), the partially reflected waves in the interior region constructively interfere, and the Fabry-Perot resonance condition is satisfied. The resonant wavenumbers obtained for a given resonator length $l$ and corrugation wavenumber $k_b$ therefore become
\ba{}\label{208}
k = \f{ (2m+1)\pi + 2\theta_1 + \alpha_1 + \alpha_2}{2l},~~m\in\mathbb{N}.
\ea 
It should be noted that $(R^{FP})^2+(T^{FP})^2=1$, which is in agreement with the principle of energy conservation. The maximum normalized free-surface elevation within the resonator is given by the so-called field enhancement parameter, i.e. (cf. the Appendix),
\ba{}\label{999}
\Xi = \f{|\mathcal{A}(x^e_1)|+|\mathcal{B}(x^e_1)|}{|\mathcal{A}(x^s_1)|} = \lp 1+R_2^B \rp \f{T^{FP}}{T_2^B}.
\ea
The highest achievable field enhancement $\Xi$ occurs when one of the Fabry-Perot resonant wavenumbers is $k=k_b/2$. Substituting $k=k_b/2$ into Eq. \eqref{208} we thus obtain the condition on the resonator length, i.e.,
\ba{}\label{209}
l \equiv l_m^B = \f{(2m+1)\pi+\theta_1+\theta_2}{k_b},~~m\in\mathbb{N}, 
\ea
leading to the highest possible $\Xi$. Interestingly, we find that Eq. \eqref{209} reduces to the classical Fabry-Perot in-phase resonance condition, that is $l=m\pi/k$, when $\theta_1+\theta_2=\pi$. This fact can be explained in terms of effective resonator length. In the case of positive corrugation slope next to the interior region for both Bragg reflectors, i.e., when $\theta_1=\pi$ and $\theta_2=0$, there is no phase shift incurred at the mirrors for waves coming from the inside at the Bragg frequency, and the effective resonator length is simply the distance $l$ between the two reflectors. The effective resonator length is instead $l+\lambda_b/2$ when $\theta_1=\theta_2=0$ as the upstream patch reflects waves in the interior with a $\pi$ phase shift. 

\begin{figure}
\centering\includegraphics[width =3.2in]{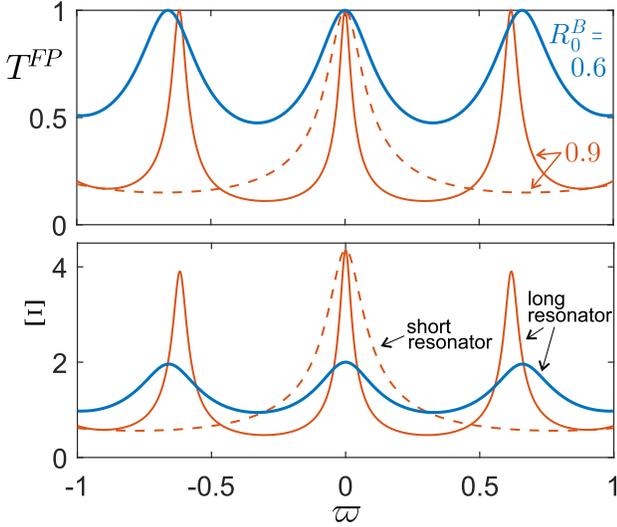}
\caption{(Color online) Transmission and field enhancement spectra for $|\varpi|\leq 1$. The number of resonant modes increases with increasing resonator length $l$ \eqref{209} ($l=l^B_{5}$: dashed line, $l=l^B_{30}$: solid lines), while the maximum field enhancement 
values $\max\Xi$ \eqref{999} increase with increasing mirror reflectivities $R^B_0=$ $R^{B}_{1,2}(\varpi=0)$. The cutoff frequency $\Omega^c$ \eqref{2012a} is the same for all three curves.  }\label{spectra}
\end{figure}

The wavelength separation $\Delta \lambda$ between adjacent transmission peaks, also called the free-spectral range (FSR) in optics \cite{Yeh1988}, can be obtained from the Fabry-Perot condition \eqref{208} as $\Delta \lambda=\lambda_{m+1}-\lambda_m$. Similar to $k$, $\Delta \lambda$ cannot in general be explicitly expressed as a function of $l$ because $\alpha_{1,2}$ are nonlinear functions of $k=2\pi/\lambda$ through $\varpi$ [cf. Eq. \eqref{2011a}]. Close to the Bragg frequency we may yet approximate it as $\Delta \lambda \approx 2\lambda_b^2/l$ (since $\alpha_{1,2}=-\theta_{1,2}$ at $\varpi=0$), showing that, as in optics, the FSR is inversely proportional to the length of the interior region. The transmission and field enhancement spectra are shown in Fig. \ref{spectra} for two different Bragg reflection coefficients (assuming $R_1^B=R_2^B$) and resonator lengths $l=l^B_m$. Comparing the two solid lines, obtained for a long interior region $l=l^B_{30}$ \eqref{209}, it is clear that large Bragg reflection coefficients result in higher transmission extinction and higher field enhancement between the two mirrors. In addition, we see that increasing $R_{1,2}^B$ modifies the FSR by changing the locations of the resonance frequencies. The FSR significantly increases when the resonator is smaller, such that the secondary Fabry-Perot resonant modes are pushed outside the
Bragg reflection bandwidth (cf. dashed line in Fig. \ref{spectra}). \\

\begin{figure}
\centering\includegraphics[width =3.2in]{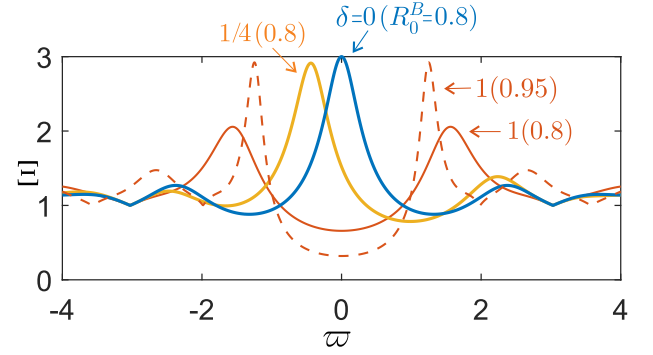}
\caption{(Color online) Field enhancement spectrum in a very short resonator of length $l=l_0^B+\delta\pi/k_b$ [cf. Eq. \eqref{209}] for $R^B_0=R^B_{1,2}(\varpi=0)=0.8,0.95$ (solid, dashed lines). $\Xi$ is maximum at the Bragg frequency when the resonator length is chosen to produce constructive interference at $\varpi=0$ ($\delta=0$), whereas the resonant peak is  shifted to lower frequencies when the interior region becomes longer ($\delta=1/4$). When $\delta=1$, the resonant modes fall outside the Bragg frequency bandwidth such that destructive interference dominate for $|\varpi|\leq 1$. The cutoff frequency $\Omega^c$ \eqref{2012a} is the same for all 4 curves. }\label{suppression}
\end{figure}

As opposed to being enhanced, incident waves whose frequencies lie within the Bragg reflection bandwidth can be suppressed between the two mirrors by detuning the resonator from the Fabry-Perot resonance condition \eqref{209}. To make this apparent, we rewrite the resonator length as
\ba{}\label{211}
l = l_m^B + \delta \pi/k_b ,
\ea
where $l_m^B$ satisfies Eq. \eqref{209}, $m\in\mathbb{N}$ and $\delta\in [0,2]$. The partially reflected waves in the interior region are in phase at the Bragg frequency, thus enhanced, when $l=l_m^B$. We show the effect of the offset parameter $\delta\neq 0$ on the field enhancement experienced by a very small resonator ($m=0$) in Fig. \ref{suppression}. When $\delta=1/4> 0$ the primary Fabry-Perot resonant mode is shifted to smaller frequencies, i.e., longer waves, because of the increased resonator length. At the critical offset $\delta=1$, the Fabry-Perot transmission peaks all lie outside the Bragg frequency bandwidth, which is therefore centered on a region with small field enhancement. The suppression strength, i.e., $\Xi^{-1}$, increases with the mirrors' reflection coefficient (cf. dashed line in Fig. \ref{suppression}). We note that while a Bragg reflection coefficient of about $80\%$ is needed to achieve an amplification $\Xi(l=l^B_m)=3$ at $\varpi=0$ ($m$ arbitrary), an equivalent field suppression of $\Xi(l=l^B_m+\pi/k_b)=1/3$ would require $|R^B_{1,2}|=95\%$, i.e., higher reflectivity mirrors. This result can be generalized analytically in the limit where $|R^B_{1,2}|\rightarrow 1$ for which $\Xi(l=l^B_m)\times \Xi(l=l^B_m+\pi/k_b)\rightarrow 4$. \\

\begin{figure}
\centering\includegraphics[width =3.4in]{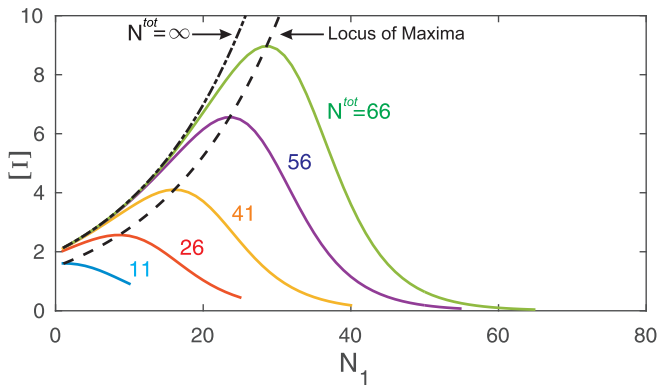}
\caption{(Color online) Significance of the ripples distributions ($N_1$ and $N_2$) on $\Xi$ for perfectly tuned waves coming from $x=-\infty$ and for various $N^{tot}=N_1+N_2$ (cf. Fig. \ref{setup}). The amplification reached at optimal distribution [cf. Eq. \eqref{000}] is shown by the dashed line, and is less than that obtained with a wall substituted for the second patch (i.e., for $N_1=N^{tot}$ and $N_2\rightarrow\infty$; see dash-dotted curve); $k_bh=1.64$, $k_b d=0.164$.}\label{dist}
\end{figure}

When waves come from all directions, the highest averaged wave amplification is achieved by setting $R^B_{1}=R^B_{2}$, or equivalently by fixing $N_1=N_2=N^{tot}/2$ where $N^{tot}$ is the total number of ripples. 
In the case where incident waves come primarily from  one direction (say, upstream of patch no. 1), however, the field enhancement within the interior region can be optimized by finding the appropriate distribution of ripples $N_{1,2}$ such that $\Xi$ is maximum for a fixed $N^{tot}$. The effect of $N_1$ on $\Xi$ is shown in Fig. \ref{dist} for perfectly tuned surface waves ($\varpi=0$), an optimal spacing [cf. Eq. \eqref{209}], and for various total number of ripples $N^{tot}$. The maximum field enhancement $\Xi$ is always obtained for $N_1^{opt}<N^{tot}/2$, i.e., $R^B_1<R^B_2$, with the difference $(N^{tot}/2-N_1^{opt})/(N^{tot}/2)$ being the greatest for low reflectivity mirrors. The optimal $N_1^{opt}$ is unique and can be obtained by maximizing $\Xi$ as a function of $N_1$ for fixed $N^{tot}$. We find
\ba{}\label{000}
N_1^{opt} = \f{1}{2s} \arctanh \left[ \f{1-2\tanh N^{tot} s}{\tanh sN^{tot} -2} \right],
\ea
where $s=\Omega^c \lambda_b/C_g$ [cf. Eqs. \eqref{2012}]. The growth of  maximum wave amplitude between the two patches is exponential with the number of corrugations since $\Xi(N_1^{opt}) \sim 2 \exp sN^{tot}/2$ when $sN^{tot}\rightarrow \infty$. Interestingly, we note that while choosing $N_1 \neq N_2$ results in a different field enhancement $\Xi$ for left-going and right-going waves, the Fabry-P\'erot reflection and transmission coefficients remain the same for both incident wave directions. This symmetry can be clearly seen from the formulas \eqref{204} for $R_{FP}$ and $T_{FP}$, which are unchanged under $R_1^B\leftrightarrow R_2^B$  swaps.   \\ 

\begin{figure}
\hspace{-0.2in}\includegraphics[width =3.3in]{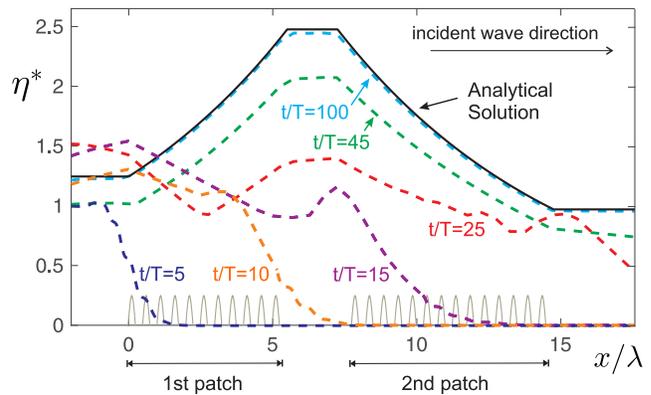}
\caption{(Color online) Transient build up of Fabry-Perot resonance for perfectly tuned water waves coming from $x=-\infty$. The resonator length is $l=l_5^*$ \eqref{209}. The normalized wave envelope $\eta^*=(|\mathcal{A}|+|\mathcal{B}|)/a_0$, with $a_0=10^{-5}h$ the incident wave amplitude, is shown at six different successive times. The physical parameters are $k_bh=1.64$, $k_bd=0.164$; $N_1=11$, $N_2=15$. Note that the amplitude of the ripples has been exaggerated in this figure. The spatial resolution of the HOS numerical simulation is $N_X=2^{12}$, and $N_T=64$ time steps were used per wave period simulated. }\label{timehis}
\end{figure}

The transient build up of wave trapping within a Fabry-Perot resonator is finally shown in Fig. \ref{timehis} for perfectly tuned waves coming from $x=-\infty$. The interior region is designed to amplify almost optimally right-going Bragg frequency waves with $k_bh=1.64$, $k_bd=0.164$, $N_1=11$, and $N_2=15$, such that $R^B_1= 60\%$ and $R^B_2=73\%$ at $\varpi=0$. Assuming a peak wave period of $7$ s, it follows that the resonator is designed to trap $52$ m long waves in $6.7$ m water depth. The numerical results are obtained utilizing the high-order-spectral (HOS) method \cite{Alam2009a,Dommermuth1987}. Trapping occurs very rapidly as shown by the beginning of increased wave envelope amplitude in the interior region for $t/T=25$. The steady state is reached after $\sim 100$ peak wave periods, which corresponds to $\sim 1.6$ times the round trip propagation time between the two most distant corrugations. \\

\section{Conclusions}

In summary, we showed that water wave trapping occurs within a water resonator made of two distinct sets of seabed corrugations, and we demonstrated the analogy with the Fabry-Perot resonance in optics. We found that the highest possible wave amplification or suppression takes place at the Bragg frequency, and we obtained the corresponding resonance condition for the resonator length \eqref{209}. Neglecting viscosity, we found that the field enhancement \eqref{999} increases infinitely with increasing mirrors' reflectivity within the validity of the linear potential flow theory. As for the Bragg reflection coefficient \eqref{2011a}, the field enhancement or suppression becomes independent of the normalized wavelength $k_bh$ in the long-wave regime, in which case the water resonator becomes fully analogous to the classical optical Fabry-Perot cavity due to the absence of dispersion by the fluid medium.  \\   

Fabry-Perot resonance of water waves may be utilized, through engineered seabed bars, to enhance wave energy extraction efficiency or to protect offshore structures. While investigation of the possibility of occurrence and the role of such effects in the dynamics of oceans is beyond the scope of this paper, we would like to comment that  the assumption of similar ripple wavelengths for the two patches is not unrealistic for naturally occurring ripples since the periodicity of, e.g., sandbars is directly dependent on the local wave conditions, which do not change much on distances of the order of a few hundreds surface wavelengths. Furthermore, the special case of a single patch of ripples adjacent to a reflecting wall can be treated similarly to a two-patch system (cf. Fig. \ref{dist}, dash-dotted line corresponding to $N^{tot}=\infty$), and may be easily realized in real oceans, e.g., in the area between nearshore sandbars and the shoreline \cite{Yu2000}. The localized amplifications of water waves excited from within a closed-ends tank with one patch of ripples \cite{Weidman2015} can also be discussed based on our analysis of the Fabry-P\'erot resonance and is in fact reminiscent of the working principle behind laser resonators.

 In either case of engineered or natural seabed bars, the Fabry-Perot resonance is expected to be important only in relatively shallow waters since the reflectivity \eqref{2011a} of Bragg mirrors decreases with increasing $k_bh$. As the water depth decreases and due to refraction, wave rays asymptotically become parallel to each other. Hence the effect of the spreading angle or multidirectionality of waves is usually neglected in such analyses \cite[e.g., Ref.][]{Davies1982} and is not pursued here. Nevertheless, we would like to comment that effects of multidirectional waves can be easily taken into account invoking the same formulation presented here and by considering an effective wavelength, which is the surface wavelength component perpendicular to the corrugation crests \cite[][]{Mei1988}. Real seabeds, particularly near shorelines, may also have a mean slope \cite{Mei2001}. In such a case and if the slope is mild, water-wave trapping is optimized by considering a slowly varying ripple wavelength, i.e., $k_b=k_b(\epsilon x)$ ($\epsilon\ll 1$), such that $\omega_b=k_bg\tanh k_bh$ remains constant everywhere \cite{Mei1988,Alam2012}. 

 The Fabry-Perot resonance of water waves is a leading order phenomenon. Therefore, even if the seabed undulations, whether engineered or natural, are not perfectly sinusoidal (due to, e.g., erosion over time or biofouling), a strong amplification or damping is achieved as long as the dominant Fourier component of the seabed satisfies the resonance condition \cite[][]{Saylor1970,Short1975,Mei2001}. In such cases, the results presented here can therefore be expected to obtain with quantitative changes proportional to the amplitudes of the non-dominant secondary topographic modes \cite[][]{Mattioli1991,Guazzelli1992}. Clearly a purely random topography does not lead to any resonance but rather result in localized waves damped because of wave energy spreading in all spatial directions \cite[][]{Alam2007}. For arbitrary corrugation shapes, large ripple amplitudes, and away from the Bragg frequency, i.e., $|\omega-\omega_b|/\omega_b \sim O(1)$, the use of Floquet theory \cite{Yu2012a,Yu2012} or numerical simulations of higher-order equations \cite{Agnon1999,Ruban2004} becomes necessary to carefully asses the degradation of the quality of the resonator \cite[][]{Guazzelli1992,Yu2010}. 

 Viscous dissipation in water, except for very short waves such as capillary-gravity waves,  is generally confined near the seabed. Due to the no-slip boundary condition at the bottom, a viscous boundary layer forms, allowing for sediment transport while dissipating wave energy. Bottom friction affects both the wave amplitude and phase \cite{Liu1986}. Following earlier studies on Bragg scattering \cite{Kirby1993}, we can infer that the water viscosity $\nu\approx 10^{-6} $ m$^2$s$^{-1}$ within a laminar boundary layer on a flat seabed results in a phase shift accumulation and amplitude attenuation rate given by $\exp[ -x(1-i)\sigma/(2C_g)]$ where $\sigma = gk^2\sqrt{\nu/(2\omega)}/(\omega\cosh^2 kh)$. For the parameters of the numerical simulation presented in Fig. \ref{timehis} and $T=7$ s, we find that $\sigma\approx 6.6~ 10^{-5} $ s$^{-1}$, which corresponds to a phase shift accumulation and wave attenuation rate of $ 0.03\%$ per wavelength. Implementing these viscous effects into our formulation we find that the field enhancement obtained in Fig. \ref{timehis} is decreased by $0.6\%$ for $l=l^B_5$ and by $3.7\%$ for $l=l^B_{100}$. Viscosity can therefore be safely neglected for a rigid smooth seabed and a resonator with an interior region of length $l\leq l^B_{100}$ since the wave field remains strongly enhanced. Indeed, for such small interior regions, the viscous phase shift is much smaller than the full-width half maximum of the field enhancement spectrum peaks. A more thorough analysis of bottom friction may, however, be necessary for erodible beds made of, e.g., sand grains since these typically exhibit stronger, though still small, viscous effects \cite[][]{Benjamin1987,Rey1995}. 

While outside the scope of the present work, we finally note that nonlinear effects, which have been shown to produce soliton-like structures over seabed corrugations \cite{Ruban2008}, could become of significance for the Fabry-Perot resonance of finite-amplitude water waves as they would most certainly limit the maximum achievable field enhancement.  

\begin{acknowledgments}
The authors wish to thank Caroline Delaire and Farid Karimpour for careful reading of the manuscript as well as Christopher Lalau-Keraly for stimulating discussions on the Fabry-P\'erot resonance in optics. Support from the American Bureau of Shipping is gratefully acknowledged.
\end{acknowledgments}

\appendix*
\section{Derivation of the Fabry-P\'erot Coefficients }

Here we derive the Fabry-Perot reflection, transmission, and field enhancement coefficients as given in Eqs. \eqref{204} and \eqref{999}. Consider a pair of water Bragg reflectors with seafloor corrugations given by Eq. \eqref{1003}. Mei \cite{Mei1985} showed that the equations governing the evolution of the wave envelopes $\mathcal{A}$ and $\mathcal{B}$ [cf. Eq. \eqref{101}] over each patch of ripples at the steady-state read 
\bsa{105}
& i\Omega\mathcal{A}_j + C_g \f{\p \mathcal{A}_j}{\p x} = -{\Omega^c \e^{i\theta_j} \mathcal{B}_j,} \\
& i\Omega\mathcal{B}_j - C_g \f{\p \mathcal{B}_j}{\p x} = {\Omega^c \e^{-i\theta_j} \mathcal{A}_j,}
\esa
in the vicinity of the Bragg frequency, i.e., $\omega=\omega_b+\Omega\sim\omega_b=\omega(k_b/2)$, with $\Omega^c = (\omega k_bd)/(4\sinh 2kh)$ the cutoff frequency, and where $j=1$ or 2 depending on whether we consider the envelope variations over region 1 or 2 (see Fig. (1)). We recall that $k_b$ and $\theta_{1,2}$ are the wavenumber and phases of the seabed corrugations, $C_g$ is the group velocity, and $\Omega$ is the detuning parameter. The time variations of the wave envelopes being sinusoidal at the steady state, we expand them out of the envelope solution by rewriting $\mathcal{A}_j(x,t)$ and $\mathcal{B}_j(x,t)$ as $A_j(x)\e^{i\Omega t}$ and $B_j(x)\e^{i\Omega t}$. The general solution to Eqs. \eqref{105} over either one of the two patches (i.e., $j=1$ or 2) can be written as \cite[e.g., Ref.][]{Yu2000}
\bsa{201}
A_j(x)=A_j(x^s_j)\mathcal P_j(x),\\
B_j(x)=A_j(x^s_j)\mathcal Q_j(x),
\esa
where
\bsa{108}
\mathcal{P}_j(x) = \lb\i q C_g\cosh q y_j-\Omega\sinh q y_j\hspace{1.5cm} \right.\nn\\
\left.+{i\e^{i\theta_j}}\Omega^cU_j\sinh q y_j\rb/{\cal I}_j,\\
\mathcal{Q}_j(x) = \lcb\lb \i q C_g\cosh q y_j+\Omega\sinh q y_j\rb U_j\hspace{1cm} \right.\nn\\
 \left.+\i\Omega^c \e^{-i\theta_j} \sinh q y_j\rcb/{\cal I}_j,
\esa  
with $y_j=x^e_j-x$ and 
\ba{} \notag
 &q Cg=\sqrt{(\Omega^c)^2-\Omega^2}, ~~ U_j=\f{B_j(x_j)}{A_j(x_j)},\nn\\
&{\cal I}_j=\i qC_g\cosh qL_j-\Omega\sinh qL_j+{i\e^{i\theta_j}}\Omega^cU_j\sinh qL_j\nn.
\ea
The envelope solution \eqref{108} provides the reflection and transmission ratios $\mathcal{Q}_j(x)$ and $\mathcal{P}_j(x)$ for the wave amplitude over the upstream and downstream patch ($j=1,2$). In the middle region ``m'', the envelope amplitudes are constants, and therefore free-surface continuity requires
\ba{}\label{202}
U_1=\f{B_1(x^e_1)}{A_1(x^e_1)}=\f{B_2(x^s_2)}{A_2(x^s_2)}\e^{-2ikl}=\mathcal{R}_2^B\e^{-2ikl}.
\ea
Assuming $U_2=0$, i.e., waves come only from the upstream side, and enforcing the condition \eqref{202}, we then obtain the Fabry-P\'erot reflection and transmission coefficients $\mathcal{R}^{FP} = \mathcal{B}_1(x^s_1)/\mathcal{A}_1(x^s_1)$ and $\mathcal{T}^{FP} = \mathcal{A}_2(x^e_2)/\mathcal{A}_1(x^s_1)$ as
\bsa{203} \label{203a}
&\mathcal{R}^{FP} = \e^{i\alpha_1}\f{|\mathcal{R}^B_1|+\mathcal{R}^B_2\e^{2i\theta_1}\e^{-2ikl}\e^{i\alpha_1}}{1+\mathcal{R}_1^B\mathcal{R}_2^B\e^{2i\theta_1}\e^{-2ikl}}, \\ \label{203b}
&\mathcal{T}^{FP} = \e^{ikl}\e^{i\theta_1}\e^{i\theta_2}\f{\sqrt{1-|\mathcal{R}^B_1|^2}\sqrt{1-|\mathcal{R}^B_2|^2}}{1+\mathcal{R}_1^B\mathcal{R}_2^B\e^{2i\theta_1}\e^{-2ikl}},
\esa
where we used the fact that [cf. Eq. \eqref{2011}]
\ba{}
\mathcal{R}^B_j = \mathcal{P}_j(x^s_j),~\text{when}~U_j=0,
\ea
along with $\alpha_j=\arg(\mathcal{R}^B_j)$. From Eq. \eqref{203} and the definition of the round trip phase shift $\gamma$ \eqref{205}, we then obtain $T^{FP}=|\mathcal{T}^{FP}|$ and $R^{FP}=|\mathcal{R}^{FP}|$ as given in Eq. \eqref{204}.

The field enhancement, defined as
\ba{}
\Xi = \f{|\mathcal{A}(x^e_1)|+|\mathcal{B}(x^e_1)|}{|\mathcal{A}(x^s_1)|}, .
\ea
and given in Eq. \eqref{999}, is also readily obtained considering that
\ba{} \notag
\Xi &= \f{|\mathcal{A}(x^s_2)|}{|\mathcal{A}(x^s_1)|} + \f{|\mathcal{B}(x^s_2)|}{|\mathcal{A}(x^s_1)|} \\ \notag
&=  \lb 1 + \f{|\mathcal{B}(x^s_2)|}{|\mathcal{A}(x^s_2)|} \rb \f{|\mathcal{A}(x^e_2)|}{|\mathcal{A}(x^s_1)|}\f{|\mathcal{A}(x^s_2)|}{|\mathcal{A}(x^e_2)|} \\
&= \lp 1+R_2^B \rp \f{T^{FP}}{T_2^B}.
\ea


\bibliography{Mainbiblio}

\end{document}

%% file: defs.tex
\def\e{{\rm e}}
\def\i{{\rm i}}
\def\d{\:{\rm d}}

\usepackage{color}

\def\ba#1\ea{\begin{align}#1\end{align}}
\def\bsa#1#2\esa{\begin{subequations}\label{#1} \begin{align}#2\end{align} \end{subequations}}

\def\d{\textrm{d}}

\def\p{\partial}
\def\be{\begin{eqnarray}}
\def\ee{\end{eqnarray}}
\def\bes{\begin{subeqnarray}}
\def\ees{\end{subeqnarray}}
\def\f{\frac}
\def\lp{\left(}
\def\rp{\right)}
\def\lb{\left[}
\def\rb{\right]}
\def\lcb{\left\{}
\def\rcb{\right\}}

\def\befi{\begin{figure}}
\def\eefi{\end{figure}}

\def\bce{\begin{center}}
\def\ece{\end{center}} 
\def\nn{\nonumber}

\usepackage{color}
\definecolor{orange}{RGB}{255,127,0}